\documentstyle[a4,12pt]{article}

\begin{document}

\begin{center}

{\LARGE Semi-Teleparallel Theories of Gravitation}

\vspace{1cm}

{\large\sc Christopher Kohler}\footnote{E-mail: {\sf c.kohler@rz.uni-sb.de}}

\vspace{0.5cm}

{\small\it Fachrichtung Theoretische Physik, Universit\"at des Saarlandes,}\\
{\small\it Postfach 151150, D-66041 Saarbr\"ucken, Germany}

\end{center}

\vspace{1cm}

\begin{abstract}
A class of theories of gravitation that naturally incorporates preferred 
frames of reference is presented. The underlying space-time geometry consists 
of a partial parallelization of space-time and has properties of Riemann-Cartan 
as well as teleparallel geometry. Within this geometry, the kinematic 
quantities of preferred frames are associated with torsion fields. Using a 
variational method, it is shown in which way action functionals for this 
geometry can be constructed. For a special action the field equations are 
derived and the coupling to spinor fields is discussed.
\end{abstract}

\vspace{1cm}

\newcommand{\ma}{{\cal M}}
\newcommand{\ve}{{\bf v}}
\newcommand{\be}{{\bf b}}
\newcommand{\ee}{{\bf e}}
\newcommand{\ze}{{\bf Z}}
\newcommand{\xe}{{\bf X}}
\newcommand{\ye}{{\bf Y}}

\newcommand{\con}{\omega_{ab}}
\newcommand{\dif}{{\rm d}}
\newcommand{\Dif}{{\rm D}}
\newcommand{\cur}{{\cal C}}
\newcommand{\om}{\mbox{\boldmath$\omega$}}
\newcommand{\gam}{\mbox{\boldmath$\gamma$}}

\newtheorem{theorem}{Theorem}
\newtheorem{definition}{Definition}

%-------------------------------------------------------------------------------

\section{Introduction}

Space-time geometries with preferred frames of reference play an important role 
in the study of the gravitational field and its quantization. Preferred frames 
are often introduced in the form of preferred coordinate systems in order to 
simplify calculations. Such coordinate systems are, for example, Gaussian and 
comoving coordinates \cite{lan}. Closely related to preferred coordinates is the 
concept of a reference medium which is used in various forms in the literature 
(\cite{bro1} and references therein). Furthermore, it was even attempted to 
introduce material reference frames that couple to gravity \cite{bro2}.

In this article, we treat preferred frames of reference in a pure geometric way 
within non-Riemannian geometry. In particular, we propose a geometrical 
formulation of the dynamics of a preferred frame. This leads to a new class of 
theories of gravitation which can, in the classification of alternatives of 
general relativity, be placed between the Einstein-Cartan theory and 
teleparallel theories of gravitation. The underlying geometry consists of a 
partial parallelization of space-time associated with a preferred frame and 
will be referred to as {\em semi-teleparallel geometry}.

The introduction of this geometry is based on a consideration of the local 
space-time symmetries associated with preferred frames. We first define what 
is understood --- in this article --- by a preferred frame of reference within 
a metric space-time geometry: The primary part is a preferred timelike vector 
field which can be normalized and represents the tangent vectors of preferred 
worldlines. Along these worldlines, we consider preferred spatial triads 
orthogonal to the worldlines. The propagation of these triads in time should 
be determined dynamically. However, since space is assumed to be locally 
isotropic, the overall orientation of the triads along the worldlines is 
irrelevant from which follows that the preferred triads are given up to a 
constant rotation on each wordline. A preferred frame can be considered to be a 
reference fluid which possesses spin degrees of freedom. In terms of local 
symmetries, the situation is as follows: The existence of a preferred timelike 
vector field breaks the local Lorentz invariance down to a local $SO(3)$ 
symmetry of spatial rotations. The existence of preferred triads on each 
worldline breaks this symmetry further down to a global $SO(3)$ symmetry on 
each worldline. Our aim is to formulate theories of gravitation that inherently 
possess this symmetry.

As a guiding principle in the formulation of theories of gravitation that 
incorporate preferred frames we assume that the preferred frame is treated 
on an equal footing with the metric tensor. At the kinematical level, this 
means that the space-time connection parallelizes the preferred timelike vector 
field throughout space-time as well as the preferred triad along each worldline 
in the same way as a metric compatible connection parallelizes the metric 
tensor. At the dynamical level, it means that the preferred frames are 
dynamical variables like the metric tensor.

In section 2 of this article, we introduce the concept of a semi-teleparallel 
geometry in detail and describe its most important properties. Similar 
geometries have been considered previously in the context of a formulation 
of a spatial tensor analysis in general relativity \cite{cat1}-\cite{fer}.
In this article, we consider the differential geometries associated with 
preferred frames from the viewpoint of a symmetry breaking and develop a 
dynamics for this type of geometry. Accordingly, our presentation is different 
from the earlier approaches. In particular, we reveal the differential 
geometric foundation of a semi-teleparallel geometry in terms of fibre bundles.

The temporal part of the space-time torsion of a semi-teleparallel geometry is 
completely determined by the kinematic quantities of the preferred frame. 
This relation is outlined in section 3.

In section 4, we consider the formulation of a dynamics for a semi-teleparallel 
geometry. We describe a general method for obtaining action functionals for 
geometries of Riemann-Cartan type. Using this method, we propose a special 
action for the semi-teleparallel geometry.

Section 5 contains a discussion of the field equations and it is shown in which 
way matter fields can be coupled to a semi-teleparallel theory of gravitation.

In section 6, we make some comments and give an outlook on possible further 
studies.

%-------------------------------------------------------------------------------

\section{Semi-Teleparallel Geometry}

We assume space-time to be a four-dimensional differentiable manifold $\ma$ 
endowed with a Lorentzian metric ${\bf g}$ of signature $(-++\,+)$. In the 
following, we suppose that on $\ma$ there is given a congruence of timelike 
worldlines being the integral curves of the tangent vector field ${\bf t}$.
The existence of the worldlines requires the topology of $\ma$ to be 
$\Sigma\times{\rm\bf R}$ where $\Sigma$ is a three-manifold.  
Rather than ${\bf t}$, we shall use the normalized vector field ${\bf v}\equiv 
\left( -g({\bf t},{\bf t})\right)^{-\frac{1}{2}} {\bf t}$ which represents the 
tangent vectors of the worldlines parameterized by their proper times. The 
vector field $\ve$ satisfies $g({\bf v},{\bf v}) = v^av_a = -1$ where ${\bf v}= 
v^a {\bf e}_a$ is written with respect to an orthonormal basis $\ee_a$ 
$( a=0,1,2,3 )$ of the tangent space $T_p{\cal M}$ at a point $p$ of $\ma$.  
Indices are raised and lowered with the Minkowski metric $\eta_{ab} = 
{\rm diag}(-1,1,1,1)$ or its inverse, that is, $v_a = \eta_{ab} v^b$.

The vector field $\ve$ defines a distribution on space-time in that it singles 
out a three-dimensional subspace of the tangent space at each point consisting 
of tangent vectors orthogonal to $\ve$. This distribution is in general not 
integrable, that is, there does in general not exist a foliation of $\ma$ by 
spacelike hypersurfaces orthogonal to $\ve$. The projection tensor $h^a_b 
\equiv \delta^a_b + v^a v_b$ can be used to project tensors onto the 
distribution.

We assume that Lorentz connections are defined on $\ma$ which are given by 
their connection 1-forms $\con$. A connection can also be introduced by the 
covariant differential ${\bf D}$ of tangent vectors ${\bf X}$, $\Dif X^a 
\equiv \dif X^a + \omega^a{}_b X^b$. ${\bf D}$ corresponds to a metric 
compatible connection which implies $\omega_{ab} = -\omega_{ba}$. We first 
consider  connections $\stackrel{G}{\omega}_{ab}$ {\it adapted to the vector 
field} $\ve$ which are defined by the condition
\begin{equation}\label{2.1}
	\stackrel{G}{\Dif}\!\!\!\:v^a \equiv \dif v^a 
	+ \stackrel{G}{\omega}{\!\!\!\;}^a{}_b v^b = 0.
\end{equation}
Eq.\ (\ref{2.1}) says that the vector field $\ve$ is a parallel vector field 
with respect to the connection $\stackrel{G}{\omega}_{ab}$. This implies a 
partial parallelization of the manifold $\ma$ in that the projection of a 
vector onto the preferred worldlines remains parallel to the worldlines after 
parallel transport along arbitrary curves on $\ma$ while the components 
orthogonal to the worldlines may be rotated. It should be remarked that the 
connection $\stackrel{G}{\omega}_{ab}$ is determined by the vector field 
$\ve$ only up to a 1-form $B_{ab}=-B_{ba}$ that satisfies $B_{ab}v^b = 0$.

In order to understand the significance of connections adapted to $\ve$, 
we establish the following result:
\begin{theorem}
	Let $\ve$ be a vector field on $\ma$ with $v^a v_a=-1$. Every metric 
compatible connection $\con$ on $\ma$ can be uniquely decomposed as follows:
\begin{equation}\label{2.2}
	\con \; = \;\;\stackrel{G}{\omega}_{ab} + \, G_{ab} ,	
\end{equation}
where $\stackrel{G}{\omega}_{ab}$ is a connection adapted to $\ve$ and $G_{ab}
= -G_{ba}$ is a 1-form satisfying $h_a^c h_b^d G_{cd} = 0$.
\end{theorem}
{\sc Proof.} Suppose that $\stackrel{G}{\omega}_{ab}$ is a
connection adapted to $\ve$. From Eq.\ (\ref{2.2}) follows
\begin{equation}\label{2.2a}
	\Dif v_a  = \;\;\stackrel{G}{\Dif}\!\!\!\:v_a 
	+ G_{ab} v^b = G_{ab} v^b.
\end{equation}
Since every antisymmetric tensor can be decomposed as
\begin{equation}\label{2.2b}
	G_{ab} = 2 v_{[a} G_{b]c} v^c + h_a^c h_b^d G_{cd},
\end{equation}
where the square brackets denote antisymmetrization, we obtain from Eq.\ 
(\ref{2.2a}), using $h_a^c h_b^d G_{cd} = 0$,
\begin{equation}\label{2.2c}
	G_{ab} = v_a \Dif v_b - v_b \Dif v_a .
\end{equation}
Hence, given a Lorentz connection $\con$, the connection 
$\stackrel{G}{\omega}_{ab}$ is uniquely given by
\begin{equation}\label{2.3}
	\stackrel{G}{\omega}_{ab}\; = \;\;\con -v_a \Dif v_b + v_b \Dif v_a .
\end{equation}
It can be verified directly that this connection is adapted to $\ve$, that is, 
satisfies Eq.\ (\ref{2.1}).

\hspace{\fill}$\Box$

\bigskip

In the case that the connection $\con$ is the torsionfree Levi-Civit\`a 
connection $\stackrel{\circ}{\omega}_{ab}$ of the metric ${\bf g}$, the 
connection $\stackrel{G}{\omega}_{ab}$ defined by Eq.\ (\ref{2.3}) can be 
physically interpreted as follows. Suppose that $\cur$ is a curve belonging to 
the congruence of preferred worldlines parameterized by the proper time $s$. 
Then Eq.\ (\ref{2.3}) specifies a parallel transport of a vector ${\bf X}$ 
along $\cur$ given by the vanishing of the derivative
\begin{equation}\label{2.4}
	\frac{\stackrel{F}{\Dif}\! X^a}{\dif s} 
	= \frac{\stackrel{\circ}{\Dif}\! X^a}{\dif s} 
	- v^a \frac{\stackrel{\circ}{\Dif}\! v_b}{\dif s} X^b
	+ v_b \frac{\stackrel{\circ}{\Dif}\! v^a}{\dif s} X^b ,
\end{equation}
where $\frac{\stackrel{\circ}{\Dif}}{\dif s}$ is the covariant derivative 
defined by the Levi-Civit\`a connection $\stackrel{\circ}{\omega}_{ab}$. This 
parallel transport is the well known Fermi-Walker transport along $\cur$ and 
Eq.\ (\ref{2.4}) is known as the Fermi derivative \cite{haw}. Eq.\ (\ref{2.3}) 
generalizes the Fermi-Walker transport in that torsion is allowed for and the 
transport may be performed in an arbitrary direction. 

Theorem 1 shows that there is a one-to-one correspondence between the set of 
Lorentz connections $\{ \con \}$ and the set of pairs 
$\{ \stackrel{G}{\omega}_{ab}, G_{ab} \}$ consisting of a connection 
$\stackrel{G}{\omega}_{ab}$ adapted to $\ve$ and a 1-form $G_{ab}$ with 
$h_a^c h_b^d G_{cd} = 0$. It should be remarked that a connection adapted to 
$\ve$ in general possesses torsion even if the associated Lorentz connection 
$\con$ is torsionfree. 

\bigskip

With the help of theorem 1 we next reveal the differential geometric origin of 
connections adapted to $\ve$ (see also \cite{cia}). For that, we identify the 
vector field $\ve$ with the basis vector ${\bf e}_0$, that is, $\ve$ has the 
fixed components $v^a = (1,0,0,0)$. The remaining basis vectors ${\bf e}_i$ 
($i=1,2,3$) at all points of $\ma$ then form a principal fibre bundle $P(\ma )$ 
over $\ma$ with structure group $SO(3)$. $P(\ma )$ is a reduced subbundle of 
the bundle of orthonormal frames $O(\ma )$ over $\ma$ with structure group 
$SO(3,1)$. The reduction is defined by the vector field ${\bf e}_0$ which 
represents a section of the associated fibre bundle $E$ over $\ma$ with 
structure group $SO(3,1)$ and with the coset $SO(3,1)/SO(3)$ as standard fibre 
\cite{kob}.

A connection on $O(\ma )$ is given by an $so(3,1)$-valued 1-form $ \om = 
\frac{1}{2}\con J^{ab}$ on $\ma$ where $J^{ab}$ are the generators of $SO(3,1)$ 
satisfying the commutation relations
\begin{equation}\label{2.5a}
	[ J_{ab}, J_{cd} ] = 2\eta_{c[a} J_{b]d} - 2\eta_{d[a} J_{b]c}.
\end{equation}
According to theorem 1, we can decompose $\om$ as 
\begin{eqnarray}\nonumber
	\om & = & \frac{1}{2}\stackrel{G}{\omega}_{ab} J^{ab} + 				
	\frac{1}{2}\left( v_a \Dif v_b -
	v_b \Dif v_a \right) J^{ab} \\ \label{2.5}
	& = & \frac{1}{2}\stackrel{G}{\omega}_{ab} J^{ab} + \Dif v_i J^{i0} .
\end{eqnarray}
Since $\Dif v_i = \omega_{i0}$ in the chosen basis, it follows from Eq.\ 
(\ref{2.5}) that $\stackrel{G}{\omega}_{i0}$ vanishes. Thus we obtain
\begin{equation}\label{2.6}
	\om = \frac{1}{2}\stackrel{G}{\omega}_{ij} J^{ij} + \Dif v_i J^{i0} .
\end{equation}
If we require $\Dif v_a = 0 $, which means that $\con$ is adapted to $\ve$, 
Eq.\ (\ref{2.6}) yields
\begin{equation}\label{2.7}
	\om = \frac{1}{2}\stackrel{G}{\omega}_{ij} J^{ij} .
\end{equation}
Since the generators $J_{ij}$ generate $SO(3)$, $\om$ is an $SO(3)$ connection.
We see that a connection adapted to $\ve$ is a Lorentz connection that is 
reducible to an $SO(3)$ connection.

\bigskip

The essential property of a connection adapted to $\ve$ is that it converts 
preferred worldlines into autoparallels by introducing  torsion in a specific 
way. We next generalize this connection to the case that a preferred frame, as 
defined in the introduction, is given. To do this, we introduce spatial 
reference frames in the form of triads of orthonormal vectors $\be_{(i)} = 
b^a_{(i)} \ee_a$ which are orthogonal to $\ve$. The $\be_{(i)}$ together with 
$\ve$ form an orthonormal basis of tangent vectors. The dual basis shall be 
denoted by $\{ \gam =\gamma_a \ee^a, \be^{(i)} = b^{(i)}_a \ee^a \}$ where 
$\ee^a$ is the cobasis corresponding to the basis $\ee_a$. Then, we have the 
following relations:
\begin{eqnarray}\nonumber
	& \gam (\ve ) = \gamma_a v^a =1 ,\qquad & \gam (\be_{(i)} ) 
	= \gamma_a b^a_{(i)} = 0 , \\ \label{2.8}	
	& \be^{(i)} ( \ve ) = b^{(i)}_a v^a = 0 ,\qquad & \be^{(i)} 
	( \be_{(j)} ) = b^{(i)}_a b^a_{(j)} = \delta^i_j .
\end{eqnarray}

In the case that the triad $\be_{(i)}$ is Fermi-Walker transported along a 
worldline, the vectors $\be_{(i)}$ on the worldline are parallel with respect 
to the corresponding connection adapted to $\ve$. If, however, there is an 
angular velocity of the triad, we can still consider the vectors $\be_{(i)}$ to 
be parallel at different points on the worldline if we modify the connection 
adapted to $\ve$ by introducing torsion appropriately. We extend the condition 
(\ref{2.1}) in the following way.
\begin{definition}
Let $\{ \ve ,\be_{(i)}\}$ be a preferred frame of reference on $\ma$. 
A connection $\stackrel{S}{\omega}_{ab}$ is called {\em semi-teleparallel} 
if it satisfies the conditions
\begin{equation}\label{2.9}
	\stackrel{S}{\Dif}\!\!\!\; v^a = 0 \quad \mbox{and} \quad 
	\stackrel{S}{\Dif}_\ve\!\!\!\; b^a_{(i)} = 0.
\end{equation}
\end{definition}

In order to gain insight into the nature of these connections, we formulate 
the following theorem which is analogous to theorem 1.
\begin{theorem}
Let $\{ \ve ,\be_{(i)}\}$ be an orthonormal basis on $\ma$. Every connection 
$\stackrel{G}{\omega}_{ab}$ adapted to $\ve$ can be uniquely decomposed as 
follows:
\begin{equation}\label{2.10}
	\stackrel{G}{\omega}_{ab}\; = \;\;\stackrel{S}{\omega}_{ab} + F_{ab} ,
\end{equation}
where $\stackrel{S}{\omega}_{ab}$ is a semi-teleparallel connection and the 
tensor $F_{ab}=-F_{ba}$ is a 1-form with $F_{ab}v^b = 0$ and 
$F_{ab}(\be_{(i)} ) = 0$.
\end{theorem}
{\sc Proof.} From Eq.\ (\ref{2.10}) follows with the help of the second of 
Eqs.\ (\ref{2.9})
\begin{equation}\label{2.10a}
	\stackrel{G}{\Dif}_\ve \!b^a_{(i)} = F^a{}_b (\ve ) b^b_{(i)} .
\end{equation}
Using $F_{ab}v^b = 0$ and the fact that $\{\be_{(i)},\ve \}$ forms an 
orthonormal basis, Eq.\ (\ref{2.10a}) can be solved for $F^a{}_b (\ve )$ 
yielding
\begin{equation}\label{2.10b}
	F^a{}_b (\ve ) = - b^a_{(i)} \stackrel{G}{\Dif}_\ve \! b^{(i)}_b .
\end{equation}
If we employ the condition $F_{ab}(\be_{(i)} ) = 0$, we conclude that
\begin{equation}\label{2.11}
	F^a{}_b = - b^a_{(i)} \stackrel{G}{\Dif}_\ve\! b^{(i)}_b \gam .
\end{equation}
Thus, given a connection $\stackrel{G}{\omega}_{ab}$ adapted to $\ve$ and a 
triad $\be_{(i)}$, there is a unique semi-teleparallel connection
\begin{equation}\label{2.12}
	\stackrel{S}{\omega}\!{}^a{}_b = \;\;\stackrel{G}{\omega}\!{}^a{}_b 
	+ b^a_{(i)} \stackrel{G}{\Dif}_\ve\! b^{(i)}_b \gam .
\end{equation}

\hspace{\fill} $\Box$

The transition from the connection $\stackrel{G}{\omega}_{ab}$ to the 
connection $\stackrel{S}{\omega}_{ab}$ has also a description in terms 
of fibre bundles. Given the bundle $P(\ma )$ of triads $\be_{(i)}$ considered 
above, we can associate with each preferred worldline $\cur$ a subbundle 
$Q(\cur)$ of $P(\ma )$ with structure group $SO(3)$ by restricting $P(\ma )$ to 
$\cur$. A triad $\be_{(i)}$ defines a reduction of each subbundle $Q(\cur)$ 
associated with a worldline to a fibre bundle with structure group $\{ 1\}$. 
Indeed, the triad along $\cur$ can be considered to be a section of the 
associated fibre bundle over $\cur$ with standard fibre $SO(3)$. Let $\om$ be 
the $SO(3)$ connection adapted to $\ve$ given by Eq.\ (\ref{2.7}). According 
to Eq.\ (\ref{2.12}), $\om$ can be decomposed as
\begin{equation}\label{2.13a}
	\om = \frac{1}{2} \stackrel{G}{\omega}_{ab} J^{ab} 
	= \frac{1}{2} \stackrel{S}{\omega}_{ab} J^{ab}
	- \frac{1}{2} b^a_{(k)} \stackrel{G}{\Dif}_\ve\! b^{(k)}_b \gam \, 
	J_a{}^b .
\end{equation}
Using the basis $\ee_0 = \ve$ and $\ee_i = \be_{(i)}$, which implies $b_{(k)}^a 
= \delta_k^a$ and $b_a^{(k)} = \delta_a^k$, Eq.\ (\ref{2.13a}) reads
\begin{equation}\label{2.13b}
	\om = \frac{1}{2} \stackrel{G}{\omega}_{ij} J^{ij}
	= \frac{1}{2} \stackrel{S}{\omega}_{ij} J^{ij}
	- \frac{1}{2} b^i_{(k)} \stackrel{G}{\Dif}_\ve\! b^{(k)}_j \gam\, 
	J_i{}^j .
\end{equation}
Since $b^i_{(k)} \stackrel{G}{\Dif}_\ve\! b^{(k)}_j = \;\; 
\stackrel{G}{\omega}{}^i{}_{j\ve}$, it follows from Eq.\ (\ref{2.13b}) that 
$\stackrel{S}{\omega}_{ij\ve}$ vanishes. This implies
\begin{equation}\label{2.13c}
	\om = \frac{1}{2} \stackrel{S}{\omega}_{ijk} \be^{(k)} J^{ij}
	- \frac{1}{2} b^i_{(k)} \stackrel{G}{\Dif}_\ve\! b^{(k)}_j \gam\, 
	J_i{}^j .
\end{equation}
In the case that $\stackrel{G}{\Dif}_\ve\! b^{(k)}_j = 0$, that is, the 
connection $\om$ is semi-teleparallel, we have
\begin{equation}\label{2.13d}
	\om = \frac{1}{2} \stackrel{S}{\omega}_{ijk} \be^{(k)} J^{ij}.
\end{equation}
With respect to the bundles $Q(\cur )$ over the worldlines, the induced 
connection then is
\begin{equation}\label{2.13e}
	\om = 0 ,
\end{equation}
which means that $\om$ is reducible to a $\{ 1\}$ connection.

In summary, a semi-teleparallel connection can be defined to be a Lorentz 
connection that is reducible to an $SO(3)$ connection and whose induced 
connection on each worldline is reducible to a $\{ 1\}$ connection.

\bigskip
Theorem 1 and theorem 2 can be combined in the following theorem.
\begin{theorem}
Let $\{ \ve ,\be_{(i)}\}$ be an orthonormal basis on $\ma$. Every Lorentz 
connection $\omega_{ab}$ can be uniquely decomposed as follows:
\begin{equation}\label{2.15}
	\omega_{ab} = \;\;\stackrel{S}{\omega}_{ab} + S_{ab} ,
\end{equation}
where $\stackrel{S}{\omega}_{ab}$ is a semi-teleparallel connection associated 
with $\{ \ve ,\be_{(i)}\}$ and the tensor field $S_{ab} = -S_{ba}$ is a 1-form 
with $ h^c_a h^d_b S_{cd} (\be_{(i)}) = 0$.
\end{theorem}

Combining Eqs.\ (\ref{2.3}) and (\ref{2.12}), we can compute an expression 
for a  semi-teleparallel connection $\stackrel{S}{\omega}_{ab}$ associated 
with a Lorentz connection $\omega_{ab}$:
\begin{equation}\label{2.14}
	\stackrel{S}{\omega}\!{}^a{}_b = \omega^a{}_b - v^a\Dif v_b 
	+ v_b \Dif_{\be_{(i)}}\!\!\: v^a
	\,\be^{(i)} + b^a_{(i)} \Dif_\ve b^{(i)}_b \gam.
\end{equation}

From this equation follows that $\stackrel{S}{\omega}_{ab}$ is invariant 
under rotations $\be^\prime_{(i)} = \Lambda^j_i\, \be_{(j)}$ of the triad for 
which $\partial_\ve \Lambda^i_j = 0$. Hence, the connection 
$\stackrel{S}{\omega}_{ab}$ is determined by the tetrad $\{ \ve, \be_{(i)} \}$ 
only up to $SO(3)$ transformations that are constant along the worldlines. 

In the following, we will use the cobasis $\ee^a$ as a basis of 1-forms for 
the connection 1-form, that is, we use the components $\omega_{abc} = 
\omega_{ab\mu}e^\mu_c$. Here, $\mu = 0,1,2,3$ is a coordinate index.
The special choice of basis $\{ \ee_0=\ve, \ee_i=\be_{(i)} \}$ will be 
called {\em semi-teleparallel frame}. Using this basis, the semi-teleparallel 
connection $\stackrel{S}{\omega}_{abc}$ is characterized by the vanishing of 
the following components:
\begin{equation}\label{2.16}
	\stackrel{S}{\omega}_{0ij} \; = 0, \qquad 
	\stackrel{S}{\omega}_{0i0} \; = 0, \qquad
	\stackrel{S}{\omega}_{ij0} \; = 0.
\end{equation}
The spatial components $\stackrel{S}{\omega}_{ijk}$ form an arbitrary $SO(3)$ 
connection.

%-------------------------------------------------------------------------------

\section{Kinematic Quantities}
If the preferred vector field $\ve$ is considered to be the velocity field of 
some form of matter, the semi-teleparallel geometry involves a new formulation 
of the kinematics of velocity fields. The kinematic quantities of a velocity 
field $\ve$ are defined using the covariant derivative 
$\stackrel{\circ}{\Dif}_a v_b$ with respect to the Levi-Civit\`a connection 
\cite{ell}. When a semi-teleparallel connection is given, this covariant 
derivative is, according to the first of Eqs.\ (\ref{2.9}), given by  
\begin{equation}\label{2.27}
	\stackrel{\circ}{\Dif}_a \!\! v_b = K_{bca} v^c,
\end{equation}
where $K_{abc} \equiv \frac{1}{2}\left( T_{abc} - T_{bac} -T_{cab} \right)$ is 
the contortion tensor of the semi-teleparallel connection and $T_{abc}$ is the 
torsion tensor defined by $T^a = \frac{1}{2} T^a{}_{bc}\ee^b\wedge\ee^c = 
\dif\wedge\ee^a + \stackrel{S}{\omega}{\!\!\!\:}^a{}_b \wedge \ee^b$. Eq.\ 
(\ref{2.27}) follows from the decomposition $\stackrel{S}{\omega}_{abc}\; 
= \;\;\stackrel{\circ}{\omega}_{abc} - K_{abc}$.

The acceleration $\dot{v}_a$ of $v_a$ is defined by $\dot{v}_a \equiv v^b 
\stackrel{\circ}{\Dif}_b\! v_a$. Eq.\ (\ref{2.27}) then yields
\begin{equation}\label{2.28}
	  \dot{v}_a = K_{abc} v^b v^c  .
\end{equation}
Hence, in a semi-teleparallel geometry, the acceleration is related to a part 
of the torsion tensor. This is also the case for the other kinematic quantities: 

The vorticity is defined by 
\begin{equation}\label{2.28a}
	\Omega_{ab} \equiv h^c_a h^d_b \stackrel{\circ}{\Dif}_{[d}\! v_{c]} .
\end{equation} 
With the help of Eq.\ (\ref{2.27}), we obtain
\begin{equation}\label{2.29}
	\Omega_{ab} = - K_{c[ab]} v^c - v_{[a} K_{b]cd} v^c v^d .
\end{equation}

The deformation tensor is given by
\begin{equation}\label{2.29a}
	\theta_{ab} \equiv h^c_a h^d_b \stackrel{\circ}{\Dif}_{(c}\! v_{d)} ,
\end{equation}
where round brackets denote symmetrization.
Eq.\ (\ref{2.27}) yields
\begin{equation}\label{2.30}
	\theta_{ab} =  - K_{c(ab)} v^c + v_{(a} K_{b)cd} v^c v^d .
\end{equation}
The trace of $\theta_{ab}$, that is, the expansion $\theta$, is given by
\begin{equation}\label{2.31}
	\theta = K^a{}_{ba} v^b .
\end{equation}
The (trace-free) shear tensor then is
\begin{equation}
	\sigma_{ab} = - K_{c(ab)} v^c + v_{(a} K_{b)cd} v^c v^d 
	-\frac{1}{3} h_{ab} K^d{}_{cd} v^c.
\end{equation}

If there are spin degrees of freedom, we can additionally define a spin 
rotation. We assume that spatial reference frames $\be_{(i)}$ along the 
worldlines of the velocity field are given with respect to which the spin 
vector is fixed. The angular velocity $\kappa_{ab}$ shall be given by
\begin{equation}\label{2.33}
	\stackrel{F}{\Dif}_\ve\! b^a_{(i)} \equiv \kappa^a{}_b\, b^b_{(i)} ,
\end{equation}
where $\stackrel{F}{\Dif}_\ve$ is the Fermi derivative. Using Eq.\ (\ref{2.4}), 
it follows
\begin{equation}\label{2.34}
	\kappa^a{}_b \,b^b_{(i)} = \;\;\stackrel{\circ}{\Dif}_\ve\! b^a_{(i)}
	- v^a \stackrel{\circ}{\Dif}_\ve\!\! v_b\, b^b_{(i)} .
\end{equation}
On the other hand, the second of Eqs.\ (\ref{2.9}) can be written as
\begin{equation}\label{2.35}
	\stackrel{\circ}{\Dif}_\ve\! b^a_{(i)} = K^a{}_{bc} v^c b^b_{(i)} .
\end{equation}
Inserting Eq.\ (\ref{2.35}) into Eq.\ (\ref{2.34}) and solving for 
$\kappa^a{}_b$, we obtain 
\begin{equation}\label{2.36}
	\kappa_{ab} = K_{abc} v^c - 2 v_{[a} K_{b]cd} v^c v^d  .
\end{equation}
While the vorticity represents the external, orbital part of the angular 
velocity, the quantity $\kappa_{ab}$ is of a pure internal nature and can be 
regarded as a spin rotation.

The kinematic quantities take on an especially simple form if we use a 
semi-teleparallel basis. The only nonvanishing components are
\begin{equation}\label{2.42}
		\dot{v}_i  =  K_{i00} ,  \qquad
		\Omega_{ij}  =  - K_{0[ij]} , \qquad
		\theta_{ij}  =  - K_{0(ij)} , \qquad
		\kappa_{ij}  =  K_{ij0} . 
\end{equation}

The contortion tensor of a semi-teleparallel connection can with the help of 
the kinematic quantities be decomposed as follows:
\begin{equation}\label{2.43}
	K_{abc} = {}^\perp K_{abc} + 2 \dot{v}_{[a} v_{b]} v_c 
		+ 2 v_{[a} \sigma_{b]c} + \frac{2}{3}\theta v_{[a} h_{b]c}
		- \kappa_{ab} v_c + 2 v_{[a} \Omega_{b]c} ,
\end{equation}
where ${}^\perp K_{abc} = h^d_a\, h^e_b\, h^f_c\, K_{def}$ is the spatial part 
of the contortion tensor, which is not determined by kinematical quantities. 
	
\medskip

We next consider the curvature of a semi-teleparallel connection.
From the definition of a semi-teleparallel connection follow the conditions
\begin{equation}\label{2.60}
	v^a\! \stackrel{S}{R}_{ab}\; = 0 
\end{equation}
for the curvature 2-form $\stackrel{S}{R}_{ab}\; \equiv \dif\,\wedge 
\stackrel{S}{\omega}_{ab} 
+ \stackrel{S}{\omega}_{ac}\wedge\stackrel{S}{\omega}{}\!^c{}_b$. These 
conditions can be expressed by the Levi-Civit\`a connection 
$\stackrel{\circ}{\omega}_{abc}$ and the contortion tensor which, according 
to Eq.\ ({\ref{2.43}), is related to the kinematic quantities. 
Eq.\ (\ref{2.60}) then represents 18 equations for the kinematic quantities 
which coincide with the known evolution and constraint equations.

Further conditions follow from the 1st Bianchi identitiy 
$\stackrel{S}{\Dif}\!\wedge \, T^a = \;\stackrel{S}{R}\!{}^a{}_b \wedge e^b$. 
Using Eq.\ (\ref{2.60}), its projection on the vector field $v^a$ is 
\begin{equation}\label{2.61}
	v_a \stackrel{S}{\Dif}\!\wedge T^a = 0.
\end{equation}
These four conditions are, when expressed by the kinematic quantities, the 
evolution and constraint equations for the vorticity already contained in 
Eq.\ (\ref{2.60}). 
	 
%-------------------------------------------------------------------------------

\section{Action Functional}

In order to find an action functional describing the dynamics of a 
semi-teleparallel space-time geometry, we use a variational method that can 
also be applied to other space-time geometries. This method makes use of theorem 
3, that is, of the property of semi-teleparallel connections to induce a unique 
decomposition of Lorentz connections. This is a characteristic shared also by 
other geometries, for example by Riemannian geometry, by teleparallel geometry, 
and --- in a trivial way --- even by Riemann-Cartan geometry itself. The 
general decomposition is
\begin{equation}\label{3.1}
	\omega_{abc} = \tilde{\omega}_{abc} + H_{abc} ,
\end{equation}
where $\omega_{abc}$ is a Lorentz connection, $\tilde{\omega}_{abc}$ stands for 
a special connection, and $H_{abc} = -H_{bac}$ is a tensor field which, 
depending on the connection  $\tilde{\omega}_{abc}$, can have restrictions. In 
the case that $\tilde{\omega}_{abc}$ is a semi-teleparallel connection, the 
spatial part of $H_{abc}$ is vanishing according to theorem 3. If 
$\tilde{\omega}_{abc}$ is a Riemannian connection or a teleparallel connection, 
$H_{abc}$ is unrestricted. If $\tilde{\omega}_{abc}$ is a Lorentz connection, 
$H_{abc}$ is zero.

To obtain an action for the connection $\tilde{\omega}_{abc}$, we start with an 
action $S_0$ that depends on the cobasis $e^a_\mu$, the Lorentz connection 
$\omega_{abc}$, the tensor field $H_{abc}$, and the contortion tensor 
$\tilde{K}_{abc}$ of the connection $\tilde{\omega}_{abc}$. The last two are, 
according to Eq.\ (\ref{3.1}), uniquely given by $\omega_{abc}$. In a next 
step, we replace $\omega_{abc}$ in $S_0$ by the decomposition (\ref{3.1}). 
Finally, we determine the stationary point of the action with respect to the 
tensor field $H_{abc}$. As a result, we obtain an action that only depends on 
the connection $\tilde{\omega}_{abc}$ and the cobasis $e^a_\mu$.

The natural choice for the action $S_0$ is a sum of the Einstein-Cartan action 
and functionals quadratic in $H_{abc}$ and $\tilde{K}_{abc}$. We first consider 
the Einstein-Cartan action,
\begin{equation}\label{3.2}
	S_{EC}[ e^a_\mu, \omega_{abc} ] = -\frac{1}{2G} \int \dif^4 x \sqrt{-g} 
	R( \omega_{abc}, e^a_\mu ) ,
\end{equation}
where $R(\omega_{abc}, e^a_\mu ) = R^{ab}{}_{ab}$ is the scalar curvature 
corresponding to the connection $\omega_{abc}$, $G$ is the gravitational 
constant, and $g$ is the determinant of the metric tensor. Inserting the 
decomposition (\ref{3.1}), we obtain up to a surface term
\begin{eqnarray}\nonumber
	\lefteqn{S_{EC} [ e^a_\mu, \tilde{\omega}_{abc}, H_{abc} ] 
	= -\frac{1}{2G} 
	\int \dif^4 x \sqrt{-g} \,\bigg[ R(e^a_\mu, \tilde{\omega}_{abc} )}
	\hspace{4cm}\\ 	\label{3.3}  
	& & {} + H^{abc}(H_{cba} - 2 \tilde{K}_{cba}) 
	+ H^a{}_{ca} ( H^{cb}{}_b 
	- 2\tilde{K}^{cb}{}_b ) \bigg] .
\end{eqnarray}

In this article, we do not consider the most general additional term quadratic 
in $H_{abc}$ and $\tilde{K}_{abc}$. Instead, we choose a term that is quadratic 
in the totally antisymmetric parts of $H_{abc}$ and $\tilde{K}_{abc}$. The 
particular combination in which $H_{abc}$ and $\tilde{K}_{abc}$ appear in Eq.\ 
(\ref{3.3}) suggests to use the term $H^{[abc]}( H_{[abc]} 
-2 \tilde{K}_{[abc]})$. Thus, we choose
\begin{eqnarray}\nonumber
	\lefteqn{ S_0 [e^a_\mu, \tilde{\omega}_{abc}, H_{abc} ] 
	= \tilde{S}_{EC} -\frac{1}{2G}
	\int \dif^4 x \sqrt{-g} \,\bigg[ H^{abc}(H_{cba} 
	- 2 \tilde{K}_{cba}) }
	\hspace{3cm} \\
	\label{3.4}& & {} + H^a{}_{ca} ( H^{cb}{}_b - 2\tilde{K}^{cb}{}_b ) 
	+\lambda H^{[abc]}( H_{[abc]} -2 \tilde{K}_{[abc]}) \bigg],
\end{eqnarray}
where
\[
	\tilde{S}_{EC} = -\frac{1}{2G} \int  \dif^4 x \sqrt{-g} 
	R(e^a_\mu, \tilde{\omega}_{abc} )
\]
and $\lambda$ is a parameter. In order to determine the stationary point of 
$S_0$ with respect to $H_{abc}$, we vary $S_0$ with respect to $H_{abc}$. 
The corresponding field equations can be solved algebraically for $H_{abc}$.

In the case that $\tilde{\omega}_{abc}$ is a Lorentz connection, $S_0$ leads 
to the Einstein-Cartan action since $H_{abc}=0$. For the Riemannian and the 
teleparallel geometry the stationary points of $S_0$ correspond to the standard 
actions which are usually considered: We first assume $\lambda \neq 1$. 
Variation of $S_0$ with respect to $H_{abc}$ leads to the condition $H_{abc} 
= \tilde{K}_{abc}$. In the Riemannian case, we have $\tilde{K}_{abc} = 0$ since 
$\tilde{\omega}_{abc}$ is the Levi-Civit\`a connection 
$\stackrel{\circ}{\omega}_{abc}$. Inserting $H_{abc} = 0$ into $S_0$, we obtain 
the Einstein-Hilbert action. In the teleparallel case, $\tilde{K}_{abc}$ is the 
contortion tensor of the teleparallel connection $\tilde{\omega}_{abc} = 0$ 
with respect to the teleparallel frame $e^a_\mu$. Since $\tilde{S}_{EC} = 0$, 
we obtain after insertion of $H_{abc} = \tilde{K}_{abc}$ into $S_0$ the action 
\begin{equation}\label{3.5}
	S_T = - \frac{1}{2G} \int \dif^4 x \sqrt{-g} 
	\left( \stackrel{\circ}{R} - 
	\lambda \tilde{K}{}^{[abc]} \tilde{K}_{[abc]}\right),
\end{equation} 
where $\stackrel{\circ}{R}$ is the Riemannian curvature scalar. $S_T$ is the 
known 1-parameter action for the teleparallel geometry \cite{mol}\cite{hay}. If 
$\lambda = 1$, the condition obtained by the variation of $S_0$ is $H_{a(bc)}
 = \tilde{K}_{a(bc)}$. The corresponding antisymmetric part of $H_{abc}$ is 
undetermined. However, insertion of the condition into $S_0$ yields again the 
Einstein-Hilbert action and the teleparallel action (\ref{3.5}) which thus 
are stationary points for all values of $\lambda$.

Since the semi-teleparallel geometry interpolates --- in a certain sense --- 
the Riemann-Cartan and the teleparallel geometry, we expect that $S_0$ yields 
a suitable action for this geometry. The determination of the stationary point 
is, however, more involved. As a simplification, we will work in a 
semi-teleparallel basis $e^a_\mu$. Then, the condition that the spatial part of 
$H_{abc}$ vanishes means $H_{ijk}=0$. We first consider the case $\lambda 
\neq 1$. Variation with respect to $H_{0ij}$ and $H_{ij0}$ leads to the 
conditions $H_{0ij}=\tilde{K}_{0ij}$ and $H_{ij0}=\tilde{K}_{ij0}$. As for 
the components $H_{0i0}$, we note that in the action $S_0$ they are contained 
only in the term $H_{0i0}\tilde{K}^{ij}{}_j$, that is, $S_0$ is linear in 
$H_{0i0}$. A stationary point exists if $\tilde{K}^{ij}{}_j =0$, which we will 
assume in the following. $H_{0i0}$ is then undetermined. The stationary point 
of $S_0$ is given by
\begin{eqnarray}\nonumber
	 S_{ST}[e^a_\mu, \tilde{\omega}_{abc}] = -\frac{1}{2G} \int 
	 \dif^4 x \sqrt{-g} \,\bigg[ R(e^a_\mu,\tilde{\omega}_{abc}) 
	 -\tilde{K}{}^{0ij} \tilde{K}_{0ji} -2 \tilde{K}{}^{0ij} \tilde{K}_{ji0}
	-(\tilde{K}{}^{0i}{}_i )^2  \\ \label{3.6}  {} -3\lambda 
	\tilde{K}{}^{[0ij]} \tilde{K}_{[0ij]} \bigg] . 
\end{eqnarray}
Using the relation
\begin{equation}\label{3.7}
	R(e^a_\mu,\tilde{\omega}_{abc}) =\; \stackrel{\circ}{R} 
	+ \tilde{K}{}^{abc} \tilde{K}_{cba} + \tilde{K}{}^a{}_{ca}
	\tilde{K}{}^{cb}{}_b + 2 \stackrel{\circ}{\Dif}_a \!\tilde{K}^{ba}{}_b ,
\end{equation}
obtained with the help of the decomposition $\tilde{\omega}_{abc} 
=\;\; \stackrel{\circ}{\omega}_{abc} - \tilde{K}_{abc}$,
Eq.\ (\ref{3.6}) can be written as 
\begin{equation}\label{3.8}
	S_{ST}[e^a_\mu, \tilde{\omega}_{ijk}] 
	= -\frac{1}{2G} \int \dif^4 x \sqrt{-g}
	\,\bigg( \stackrel{\circ}{R} + \tilde{K}{}^{ijk} \tilde{K}_{kji}
	-3\lambda \tilde{K}{}^{[0ij]} \tilde{K}_{[0ij]} \bigg) ,
\end{equation}
where a surface term has been omitted.
In the case $\lambda = 1$, there does not exist a unique stationary point. We 
obtain the condition $H_{0ij} - \tilde{K}_{0ij} = H_{ij0} - \tilde{K}_{ij0}$ 
which when inserted into $S_0$ leads also to the action (\ref{3.8}).

We use the action (\ref{3.8}) as an action for the semi-teleparallel geometry. 
$S_{ST}$ reflects the characteristic feature of a semi-teleparallel geometry to 
be a Riemann-Cartan geometry in the spatial part and a teleparallel  geometry 
in the temporal part: The second term in the integrand corresponds to an 
Einstein-Cartan gravitation in the spatial projection of space-time while 
the third term pertains to a teleparallel gravitation in the temporal 
projection.

%-------------------------------------------------------------------------------

\section{Field Equations and Matter Coupling}

We next derive the field equations in the matter free case following from the 
action (\ref{3.8}). Only the second term in the action depends on the spatial 
part $\omega_{ijk}$ of the connection. Through variation with respect to 
$\omega_{ijk}$ we obtain the vanishing of the spatial torsion, 
$\tilde{T}_{ijk}=0$, that is, the spatial geometry is Riemannian. Note that the 
condition $\tilde{K}^{ij}{}_j = 0$ is consistent with these field equations.

Variation with respect to the semi-teleparallel frame $e^a_\mu$ leads to 16 
field equations which, after contraction by $e^b_\mu$ and subtracting the 
trace, can be divided into the symmetric part 
\begin{eqnarray}\nonumber
	\stackrel{\circ}{{\cal R}}_{00} -\, \lambda
	K^{ij}{}_0 K_{[0ij]} = 0, \\
	\nonumber
	\stackrel{\circ}{{\cal R}}_{0k} -\, \lambda
	\left( K^i{}_{00} K_{[0ik]} +
	\frac{1}{2}\stackrel{\circ}{\omega}{}\!^{ij}{}_k
	K_{[0ij]} \right) = 0 , \\
	\label{4.1}
	\stackrel{\circ}{{\cal R}}_{kl} -\, \lambda
	\left( K^i{}_{0k} K_{[0il]} + K^i{}_{0l} K_{[0ik]} \right) = 0,
\end{eqnarray}
and the antisymmetric part
\begin{eqnarray}\nonumber
	\partial_\mu \left( \sqrt{-g}\, e^{\mu j} K_{[0jk]} \right) -\frac{3}{2}
	\sqrt{-g} \stackrel{\circ}{\omega}_{[kij]} K^{[0ij]} = 0, \\
	\label{4.2}
	\partial_\mu \left( \sqrt{-g}\, e^\mu_0 K_{[0kl]} \right) = 0 ,
\end{eqnarray}
where $\stackrel{\circ}{{\cal R}}_{ab}$ is the Ricci tensor of the 
Levi-Civit\`a connection and we have removed the overtilde of $K_{abc}$.

The temporal part of the contortion tensor appearing in
Eqs.\ (\ref{4.1}) and (\ref{4.2}) can be expressed by the kinematic quantities 
of the semi-teleparallel frame according to Eqs.\ (\ref{2.42}).
Eqs.\ (\ref{4.1}) and (\ref{4.2}), therefore, are field equations for the 
semi-teleparallel frame. They are similar to the field equations of the 
teleparallel gravitation \cite{hay}, they contain, however, only the temporal 
part of the torsion tensor. As in the case of the teleparallel gravitation, 
one expects problems with the time evolution of the semi-teleparallel frame in 
that the frame is not uniquely determined by the initial conditions, at least 
for some special solutions \cite{kop}\cite{nes}. 

Since scalar fields and gauge fields do not couple to the space-time torsion, 
the incorporation of these fields is unproblematic. We will now consider the 
incorporation of spinor fields. Simply adding a spinor action in the 
semi-teleparallel geometry to the action (\ref{3.8}) leads to inconsistencies 
since the field equations then require $\tilde{K}^{ij}{}_j \neq 0$ in general. 
Instead, we repeat the construction of the action given in the previous 
section beginning with the action $S_0$ where a spinor action in the 
Riemann-Cartan space-time is added and the decomposition (\ref{3.1}) is used. 
We limit ourselves to the case of a Dirac field. We add to Eq.\ (\ref{3.4}) the 
Dirac action
\begin{eqnarray}\nonumber
	S_D & = & \int \dif^4 x \sqrt{-g} \left[ 
	\frac{{\rm i}}{2}  \left( \bar{\psi}\,\gamma^a \nabla_{\! a} 			
	\psi - \overline{\nabla_{\! a} \psi}\, \gamma^a\psi \right) 
	- m\bar{\psi}\psi \right] \\ \label{4.3}
	& = & \tilde{S}_D
	+ \int \dif^4 x \sqrt{-g} H_{abc} \tau^{abc}   ,
\end{eqnarray}
where $\tilde{S}_D$ is the Dirac action with respect to a semi-teleparallel 
connection, $\nabla_a \equiv e^\mu_a \partial_\mu + \frac{1}{2}\omega_{bca} 
\Sigma^{bc}$ with $\Sigma^{ab} \equiv -\frac{1}{2}\gamma^{[a}\gamma^{b]}$,
and 
\begin{equation}
	\tau^{abc} \equiv \frac{{\rm i}}{4} \bar{\psi} 
	\left( \gamma^c \Sigma^{ab} + \Sigma^{ab}\gamma^c \right) \psi 
	= -\frac{{\rm i}}{4} \bar{\psi} \gamma^{[a} \gamma^b \gamma^{c]} \psi  
\end{equation}
is the totally antisymmetric spin angular momentum.
We again use a semi-teleparallel frame where $H_{ijk} =0$ and determine the 
stationary point of $S_0 + S_D$ with respect to the remaining components of 
$H_{abc}$. We obtain
\begin{equation}\label{4.4}
	H_{0ij} = \tilde{K}_{0ij} + \frac{G}{\lambda - 1} \tau_{0ij} ,\quad
	H_{ij0} = \tilde{K}_{ij0} + \frac{G}{\lambda - 1} \tau_{0ij} ,\quad
	\tilde{K}^{ij}{}_j = 0 ,
\end{equation}
where  $\lambda = 1$ is excluded. In the case $\lambda = 1$, there only exists 
a stationary point if $\tau_{0ij} = 0$ for which case the consideration of the 
previous section apply.
Inserting Eq.\ (\ref{4.4}) into the action $S_0 + S_D$, we obtain
\begin{eqnarray}\nonumber
	S_{STD} &=& - \frac{1}{2G} \int \dif^4 x \sqrt{-g} \left(
	\stackrel{\circ}{R} + \tilde{K}^{ijk}\tilde{K}_{kji} 
	-3\lambda \tilde{K}^{[0ij]} \tilde{K}_{[0ij]} \right) \\
	\nonumber
	&{}& + \int \dif^4 x \sqrt{-g} \left[ \frac{{\rm i}}{2}  
	\left( \bar{\psi}\,\gamma^a \stackrel{\circ}{\nabla}_{\! a} \psi
	- \overline{\stackrel{\circ}{\nabla}_{\! a} \psi}\, \gamma^a\psi 
	\right) - m\bar{\psi}\psi - \tilde{K}^{ijk} \tau_{ijk} - \frac{3G}{2} 					
	\frac{\tau^{0ij}\tau_{0ij}}
	{\lambda - 1} \right] \\
	\label{4.5}
	&=& S_{ST} + \stackrel{\circ}{S}_D - \int \dif^4 x \sqrt{-g} \left(
	\tilde{K}^{ijk} \tau_{ijk} - \frac{3G}{2} \frac{\tau^{0ij}\tau_{0ij}}
	{\lambda - 1} \right) ,
\end{eqnarray}
where  $\stackrel{\circ}{S}_D$ is the Dirac action in Riemannian space-time. 
Only the spatial part of the torsion couples to the spin angular momentum. 
Furthermore, we obtain a spin-spin contact interaction which contains the 
temporal part of the spin angular momentum and which depends on the parameter 
$\lambda$.

%-------------------------------------------------------------------------------

\section{Discussion}

The theory of gravitation proposed in this article relies on a space-time 
geometry that is formally a constrained Riemann-Cartan geometry and can be 
considered to be a mixture of Riemann-Cartan and teleparallel geometry. This 
semi-teleparallel geometry should, however, be seen as a separate geometry since
--- in contrast to other geometries --- it incorporates the concept of preferred 
frames of reference in a natural way. Accordingly, theories of gravitation based 
on semi-teleparallel geometry are qualitatively different from Einstein-Cartan 
and teleparallel gravitation since they provide a dynamics of preferred frames, 
which is not contained in the last two. 

We have not specified in this article a physical interpretation of the preferred 
frames. The formalism allows for several options. For example, the preferred 
vector field could be a Killing vector field. Another possible interpretation 
of the preferred frames is in terms of the rest frame of matter. In this case, 
however, one encounters large temporal torsion fields, in particular those 
associated with the shear of the velocity field. 

Although we have considered in this article the four-dimensional case, a 
semi-tele\-paral\-lel geometry can be defined on manifolds of arbitrary 
dimension. Furthermore, the congruence of worldlines can be generalized to 
congruences of extended objects of arbitrary dimension. The fibre bundle 
description of the semi-teleparallel geometry given in section 2 can be easily 
generalized and provides then a method for introducing a semi-teleparallel 
geometry on manifolds with several preferred vector fields. In this context, it 
should be noted that the Riemann-Cartan and the teleparallel geometry 
correspond to generalized semi-teleparallel geometries where the extended 
objects are the single space-time points and space-time itself, respectively. 
Physical applications for the case of higher-dimensional space-times are 
Kaluza-Klein theories the preferred frames being specified by spacelike Killing 
vectors. Since these vector fields are not dynamical, the corresponding 
semi-teleparallel theories of gravitation are simpler than the theories given 
in this article. The five-dimensional case leads to the Einstein-Cartan-Maxwell 
theory \cite{koh}.

The definition of the preferred frame of reference and of the corresponding 
semi-teleparallel geometry is restricted to metric-compatible geometries in 
this article. It is possible to extend the formalism to metric-affine 
geometries by allowing the spatial triad to undergo $SL(3,{\bf R})$ 
transformations.

Further possible studies are the time evolution of the semi-teleparallel 
theories of gravitation within the canonical formalism and the 
semi-teleparallel gravitation in $2+1$ dimensions.

%-------------------------------------------------------------------------------

%-------------------------------------------------------------------------------

\end{document}